\documentclass[a4paper,12pt,english]{article}
\pdfoutput=1

\usepackage[bindingoffset=0.3cm,textheight=22.5cm,hdivide={2.7cm,*,2.7cm}, vdivide={*,22cm,*}]{geometry}
\usepackage[bookmarksnumbered=true,breaklinks=true]{hyperref}

\usepackage{amsmath,amsfonts,amssymb,babel,slashed,graphicx,color}

\usepackage[numbers,square,comma,sort&compress]{natbib}

\makeatletter

\global\long\def\bra#1{\left\langle #1\right|}
\global\long\def\ket#1{\left|#1\right\rangle }

\newcommand{\bbm}{\left(\begin{matrix}}
\newcommand{\ebm}{\end{matrix}\right)}
\newcommand{\beq}{\begin{eqnarray}}
\newcommand{\eeq}{\end{eqnarray}}
\makeatother

\newcommand{\del}{\partial}

\newcommand{\be}{\begin{equation}}
\newcommand{\ee}{\end{equation}}

\newcommand{\beqa}{\begin{eqnarray}}
\newcommand{\eeqa}{\end{eqnarray}} \newcommand{\eq}[1]{(\ref{#1})}
\def\nn{\nonumber} \def \bea{\begin{eqnarray}} \def\eea{\end{eqnarray}}
\def\obar{\overline}
\newcommand{\barr}{\begin{array}}
\newcommand{\earr}{\end{array}}
\numberwithin{equation}{section}

\def\a{\alpha}  \def\b{\beta}
  \def\G{\Gamma}
 \def\d{\delta} \def\D{\Delta}

 \def\L{\Lambda}

\def\cA{{\cal A}} \def\cB{{\cal B}}  
  \def\cF{{\cal F}} 
 \def\cH{{\cal H}}  
   
\def\cM{{\cal M}}  \def\cO{{\cal O}} 
\def\cP{{\cal P}}

\def\R{{\mathbb R}} \def\C{{\mathbb C}} 

 \def\one{\mbox{1 \kern-.59em {\rm l}}}

\def\mmu{\mathfrak{u}}

\def\bit{\begin{itemize}} \def\eit{\end{itemize}} 
\def\Tr{{\rm Tr}}
\def\tr{{\rm tr}}

\def\({\left(} \def\){\right)}

 \allowdisplaybreaks[3]

\begin{document}

\makeatother


\parindent=0cm

\renewcommand{\title}[1]{\vspace{10mm}\noindent{\Large{\bf

#1}}\vspace{8mm}} \newcommand{\authors}[1]{\noindent{\large

#1}\vspace{5mm}} \newcommand{\address}[1]{{\itshape #1\vspace{2mm}}}


\begin{titlepage}
\begin{flushright}
 UWThPh-2016-9 
\end{flushright}
\begin{center}
\title{ {\Large String states, loops and effective actions \\[1ex]  in noncommutative field theory and matrix models} }

\vskip 3mm

\authors{Harold C. Steinacker{\footnote{harold.steinacker@univie.ac.at}}
}
 
\vskip 3mm

 \address{ 

{\it Faculty of Physics, University of Vienna\\
Boltzmanngasse 5, A-1090 Vienna, Austria  }  
  }

\bigskip

\vskip 1.4cm

\textbf{Abstract}
\vskip 3mm

\begin{minipage}{14cm}%

Refining previous work by Iso, Kawai and Kitazawa, we discuss bi-local string states as 
a  tool for loop computations in noncommutative field theory and matrix models.
Defined in terms of coherent states, they
exhibit the stringy features of noncommutative field theory.
This leads to a  closed form for the 1-loop effective action in position space,
capturing the long-range non-local UV/IR mixing for scalar fields. 
The formalism applies to generic fuzzy spaces.
The non-locality is tamed in the maximally supersymmetric IKKT or IIB model, 
where it gives rise to supergravity. The linearized supergravity interactions  are obtained 
directly in position space  at one loop using  string states on generic noncommutative branes.

\end{minipage}

\end{center}

\end{titlepage}

\tableofcontents

\section{Introduction}

Noncommutative field theory (NCFT) was conceived  as a generalization of (quantum) field theory 
to noncommutative or quantized spaces. One of the early hopes was that the intrinsic  
uncertainty scale of the geometry would lead to a UV regularization of the corresponding field theory.
However, it turned out that this is not the case. Rather, the phenomenon of UV/IR mixing
\cite{Minwalla:1999px} leads to an unexpected behavior of the quantum effective action 
at low energies, and  IR divergences arise due to  UV contributions in the loops.
This phenomenon was partially understood from various points of view, see e.g.
\cite{Liu:2000qhb,Kinar:2001yk,VanRaamsdonk:2001jd,Armoni:2001uw,Douglas:2001ba,Szabo:2001kg,Grosse:2008xr,Blaschke:2010rr} and references therein. 
The realization of noncommutative field theory in string theory \cite{Seiberg:1999vs} 
suggested an interpretation in terms of a closed string exchange \cite{Armoni:2001uw}, 
a geometric understanding in terms of emergent gravity was found  \cite{Steinacker:2007dq}, and
a relation with non-locality was exhibited \cite{Iso:2000ew,Jiang:2001qa,Kinar:2001yk,VanRaamsdonk:2001jd}.
In any case, UV/IR mixing means that noncommutative field theory is not simply 
a deformation of ordinary field theory, but is qualitatively different.

In this paper, we consider a powerful tool in the framework of noncommutative field theory
given by string states, refining and developing the ideas introduced in \cite{Iso:2000ew}. 
These states make the string-like character of NC field theory manifest,  
they provide a clear understanding of 
UV/IR mixing, and an efficient  way to compute loop integrals.
String states are defined as $\ket{x}\bra{y} \in End(\cH)$, in terms of coherent states $|x\rangle$ 
on the noncommutative space under 
consideration. They are elements of 
the noncommutative algebra  of functions on the space, but they have no classical analog
in field theory. They play a dominant role in the loop integrals, 
which  explains the stringy nature of NCFT.

One of the technical results of this paper is a representation of  
one-loop integrals on fuzzy spaces
in terms of integrals over string states rather than group-theoretical harmonics.
This was developed to find a practical way of evaluating 
loop  corrections on such backgrounds in Yang-Mills matrix models.
The standard way of evaluating these loop integrals is to use a group-theoretical basis 
of functions (such as spherical harmonics on the fuzzy sphere). However, 
this  leads to unreasonable 
difficulties, requiring the asymptotics of various group-theoretical objects such as 
6J symbols and their higher analogs. Moreover on generic spaces without symmetry, such a 
computation was practically impossible outside of the semi-classical regime.
Most importantly, the group-theoretical approach hides the physical meaning of the results.
Although the main ideas of the present approach are contained in  \cite{Iso:2000ew},
we improve their results by replacing the ad-hoc lattices by an  integration in position space, which yields
 a simple closed formulas for the effective action in position space.

We first review  the basic facts about coherent states on the fuzzy sphere,
which generalize to any quantized compact coadjoint orbit. In particular, 
the separation of the space of  function into the 
semi-classical IR regime and the - much larger - UV regime is carefully discussed. 
The latter is best described by the string states, which are 
interpreted as strings whose energy and momentum is given by their length.
These states have the remarkable property that they 
(approximately) diagonalize the Laplacian, and are ``bi-local'' in configuration space.
The corresponding propagator takes a very simple form, 
which makes them ideally suited for quantization. 
An over-completeness relation 
leads to an exact representation of the trace in the one-loop effective action.
We apply this  in the basic one-loop integrals, and  obtain a closed form 
for the (quadratic) 1-loop effective action in position space. This works for any quantized 
coadjoint orbit, and reproduces the known results for the fuzzy sphere
which were obtained originally in a  more complicated and less transparent way. 
On the Moyal-Weyl quantum plane, the origin of the non-local UV/IR mixing is clarified.
The generalization  to generic fuzzy spaces and to higher-loop computations is 
also discussed.

The results clearly exhibit the non-local nature of generic noncommutative 
field theories at the quantum level, making 
the previous observations in \cite{VanRaamsdonk:2001jd,Kinar:2001yk,Jiang:2001qa} more explicit and manifest.
Hence  attempts to directly use generic (non-supersymmetric) NC field theories 
as a replacement for ordinary local QFT are doomed\footnote{One may however consider 
various limits of noncommutative field theories, which may again become local, see e.g. \cite{Grosse:2012uv}.}, 
and only the maximally supersymmetric model(s) remain as  candidates for a fundamental, ``UV-complete'' quantum theory.

This is the subject of the second part of this paper, where 
the formalism of string states is used to elaborate the 1-loop effective action 
of the supersymmetric IKKT or IIB model. 
In this case, the residual non-locality is mild and can be understood as  a manifestation of the 
10-dimensional supergravity in target space, which leads to a short-range $r^{-8}$ interaction.
It is indeed expected  that this model is closely related to  
IIB supergravity and string theory.
Up to now, this could be verified from the matrix model side (mostly for the BFSS model)
only for  simple  configurations such as parallel 
or spherical branes or for separate objects represented by  block-matrices  
\cite{Chepelev:1997av,Kabat:1997sa,Kabat:1997im,Chepelev:1998sm,Okawa:1998pz,Douglas:1998tk,Taylor:1998tv,Taylor:2001vb,Kitazawa:2002vh},
possibly with some higher multipole moments.
However a derivation for generic (noncommutative) branes was missing and quite out of reach so far.
The present formalism allows to generalize the old arguments to a much more general setting, and 
gives explicitly the 10D supergravity interactions in position space.
This is very important in the on-going effort to analytically understand the physics of branes 
in this model, which is a candidate for a theory of fundamental interactions including gravity.

In particular, the present paper provides the necessary techniques for 1-loop computations on  
the fuzzy 4-sphere in the IKKT model. This is presented in a separate paper, demonstrating
the emergence of 4-dimensional gravity \cite{Steinacker:2016yy}.

\section{Coherent states and string states}

\subsection{Coherent states on the fuzzy sphere}
\label{coherent-fuzzy-S-2}

The fuzzy 2-sphere $S^2_N$  \cite{hoppe1982QuaTheMasRelSurTwoBouStaPro,Madore:1991bw} is defined in terms of 
3 hermitian matrices $X^a, \ a=1,2,3$
which satisfy the algebra
\begin{align}
[X^a,X^b] = i \varepsilon^{abc} X^c,
 \qquad  X^a X_a = \frac 14 (N^2-1) \ =: R_N^2\ .
 \label{fuzzy-S2-def}
\end{align}
Hence $X^a = J^a_{(N)}$  generate the irreducible representation of $SU(2)$
on $\cH = \C^N$.
 Functions on $S^2_N$ are given by (possibly hermitian) elements of the algebra
 $\cA = End(\cH)$, which decomposes as $SU(2)$-module into fuzzy spherical harmonics $\hat Y^l_m$ according to
 $\cA = \oplus_{l=0}^{N-1}\ (2l+1)$. Here $(n)$ denotes the $SO(3)$ irrep with dimension $n$.
 The matrix Laplacian is defined as
 \begin{align}
  \Box \phi = [X^a,[X_a,\phi]], \qquad \quad \phi \in End(\cH)
 \end{align}
and it is easy to see that it has the same spectrum $l(l+1)$ for $l=0,1,2,..,N-1$ as the classical 
Laplacian on the sphere, and $\hat Y^l_m$ are the eigenfunctions.
 The commutation relations \eq{fuzzy-S2-def} state that fuzzy $S^2_N$ is a quantization of $\cM=S^2$ with 
 the $SO(3)$-invariant symplectic form $\omega$ (or Poisson structure) satisfying the quantization condition 
 \begin{align}
  \int_\cM \omega = 2\pi \dim(\cH).
 \end{align}
This construction generalizes to any (quantized) coadjoint orbit $\cM$ of a compact Lie group, see e.g. 
\cite{Hawkins:1997gj,Steinacker:2011ix}.

As for all quantized coadjoint orbits, coherent states on $\cM = S^2 = SU(2)/U(1)$
are given by highest weight states $|\L\rangle\in \cH$ and their $SU(2)$ orbits \cite{Perelomov:1986tf},
\begin{align}
 |x\rangle &= g_{x} \cdot |\L\rangle, \qquad g_{x} \in \  SU(2)  \nn\\[1ex]
 x^a &=  \langle{ x}| X^a |{ x}\rangle \equiv \langle X^a\rangle \ 
 \qquad  \qquad  x^a x_a = \frac 14(N-1)^2 =: r_N^2 \ .
\end{align}
Here  $r_N^2$ is the radius of the coherent state orbit.
Up to a $U(1)$ phase factor, they are in one-to-one correspondence to 
points $x$ on $\cM$. We therefore label them locally by $x\in \cM$, where 
the ``north pole`` $p\in \cM$ corresponds to the highest weight state $|\L\rangle$.
They are optimally localized as follows
\begin{align}
\Delta^2  
  &= \sum_a\langle (X^a)^2\rangle  - \langle  X^a \rangle^2 
   = R_N^2 - r_N^2  = \frac{N-1}{2} \nn\\
  &=: L_{NC}^2  \ \ \ll R_N^2 , \qquad N \gg 1   \ .
 \label{Delta-S4}
\end{align}
$\Delta^2$ is a measure for the uncertainty in position space,
which defines the noncommutativity scale $L_{NC}$.
Upon rescaling $X \to r X$, the sphere can have any desired radius $R$, and 
$L_{NC} \sim\frac{R}{\sqrt{N}} \to 0$
as $N\to \infty$ for fixed $R$.
It is easy to see that the uncertainty is minimized for the coherent states;
for more details and illustrations see e.g. \cite{Schneiderbauer:2016wub}.
Furthermore, the coherent states $|x\rangle$ on $S^2_N$ form an over-complete basis, with 
\begin{align}
  \one_\cH &= c_N\,\int dx | x\rangle  \langle x| , \qquad c_N =  \frac{\dim \cH}{\rm Vol \cM}\ .
  \label{coherent-states-S2}
\end{align}
Indeed the operator defined on the rhs is invariant under the adjoint action 
of $SU(2)$, and the only operator with this property is $\sim \one$ 
(because $\cH$ is  irreducible).
This gives the following representation of the trace of any operator $\cO\in End(\cH)$
\begin{align}
   \tr \cO &= \frac{\dim \cH}{\rm Vol \cM}\,\int dx \langle x| \cO | x\rangle \ .
   \label{trace-coherent}
\end{align}
Here $\tr$ denotes the trace on $\cH$.
The overlap of the coherent states decays rapidly  with the distance between $x$ and $y$, 
\begin{align}
 |\langle x|y\rangle|^2 &= \frac 1{c_N }\,\d_N(x,y) \qquad \to \  0 \ \  \mbox{for} \ \ x\neq y, \ \ N \to \infty
\label{delta-N}
\end{align}
which defines a regularized delta function $\int dx \d_N(x,y) = 1$ on $\cM$. 
On the fuzzy sphere, there is an explicit formula \cite{Perelomov:1986tf}
\begin{align}
  |\langle x |y\rangle|^2  &= (\frac{1+x\cdot y}2)^{N-1} 
   \ \approx \ \exp(-\frac 14 \phi^2 (N-1)) , \qquad \phi^2 \ll 1  
\end{align}
where $\phi$ is the angle between $x$ and $y$.
Hence $\d_N(x,y)$ is localized on an area  $\frac{4\pi}{N}$, which reflects the 
quantization of the sphere in terms of $N$ quantum cells.

The phase of $\langle x|y\rangle$ also contains interesting information.
Since the coherent states are determined only up to a $U(1)$ phase, 
they form a $U(1)$ bundle $\cB$ over $S^2$.
Near some point $p\in S^2$ (the north pole, say), 
we can define a local section $|x\rangle = e^{i\phi_i J_i}|0\rangle$
parametrized by 2 angles $\phi_i, \, i=1,2$ relative to $p$.
The group action defines a connection 
$\nabla$, with curvature\footnote{this is the line bundle with 
monopole number $N-1$.} given by the symplectic form
underlying the quantum space, just like in quantum mechanics.
Then one finds
\begin{align}
  \langle x |y\rangle &= e^{iA(x,y)}(\frac{1+x\cdot y}2)^{(N-1)/2}  
  \ =: \ \frac 1{c_N}\, \tilde\d_N(x,y) 
  \label{innerproduct-delta}
\end{align}
where $\tilde\d_N(x,y)$ is again a (now complex-valued) approximate delta function which satisfies 
\begin{align}
 \int dx \,\tilde\d_N(x,y) \, |x\rangle  = |y\rangle
\end{align}
with similar localization properties.
Here $A(x,y)$ is the symplectic area of the spherical triangle spanned by $x,y, p$.

\paragraph{Operators and symbols.}

Coherent states provide a  useful and explicit link 
between functions on $\cM$ and operators.
For an arbitrary operator $\cO\in End(\cH)$, we define the {\em symbol} of $\cO$ to be 
\begin{align}
 \cO(x) = \bra{x} \cO \ket{x} \ .
 \label{symbol}
\end{align}
This should be viewed as de-quantization of $\cO \sim \cO(x)$.
In particular, 
 $x^a = \bra{x} X^a \ket{x}$. 
Combining this with \eq{trace-coherent}, we can write the trace in the familiar form
\begin{align}
  \tr \cO &=  c_N\,\int dx\, \cO(x) \ .
\end{align}
Conversely, one can certainly represent every fuzzy function as 
\begin{align}
 \cO =  c_N^2\int dx dy \langle x| \cO |y\rangle  |x\rangle\langle y| 
\end{align}
however this is far from unique. At least on quantized homogeneous spaces 
one can even find a diagonal representation
\begin{align}
 \cO =  c_N\int  dx  \tilde O(x)  |x\rangle\langle x| \ ,
\end{align}
however $\tilde O(x) \neq \cO(x)$ in general.
For example on $S^2_N$, we can write
\begin{align}
 \hat Y^l_m = c_N\int_{S^2} d x  Y^l_m(x)  |x\rangle\langle x|
\end{align}
because both sides transform in the same way under $SO(3)$.
Similarly, plane waves on the Moyal-Weyl plane $\R^n_\theta$ can be written as  
\begin{align}
 e^{i k X}  = c \int d x e^{ik x} |x\rangle\langle x| \ .
\end{align}
Hence all functions on fuzzy spaces can be represented in this diagonal way, however this 
 is very delicate for large momenta and may be completely 
misleading\footnote{in the same vein, using a star product 
for loop computations in NC field theory is  misleading.} as we will see. 
It should only be used in the semi-classical low-energy sector, which is defined as follows:

\paragraph{IR sector.}

The important property which characterizes the semi-classical or low energy regime for
functions on fuzzy spaces is their approximate {\em locality}. 
An operator or fuzzy function is in the semi-classical low energy (IR)  regime
if the {\em non-local} matrix elements decay at distance scales $|x-y| \sim L_{NC}$, so that
\begin{align}
 \langle x| \cO |y\rangle \approx  \langle x| \cO |x\rangle \ \d_N(x,y) \ .
\end{align}
This is the crucial property for external fields which will justify the following methods 
for computing the effective action. In particular, we will need 
\begin{align}
 \bra{x} f(X) \ket{y} \approx f(x) \langle x|y\rangle \approx f(y) \langle x|y\rangle \ ,
\end{align}
which holds for  functions $f(x)$ which are approximately constant 
on the scale $L_{NC}$.
For (low) polynomials in $X^a$, this
follows from the fact that the dispersion of $X^a$ is given by  $\Delta^2$,
which is precisely the NC scale.
The maximal angular momentum  compatible with this requirement 
is $l \leq \sqrt{N}$, which
 is  precisely the  scale of the fuzzy delta function localized e.g.
at the north pole,
\begin{align}
  |p\rangle\langle p| &=: \  \frac 1{c_N}\,\d_N(X;p) \ 
\end{align}
with symbol $\frac 1{c_N }\,\d_N(x,y)$ \eq{delta-N}.
This is  optimally localized with 
uncertainty $\D^2 \sim L_{NC}^2$ \eq{Delta-S4},
and  has angular momentum $l_{NC} \sim \|[X_i,.]\| \sim \sqrt{N}$.

\paragraph{UV sector.}

In contrast, most of the operators $\cO\in End(\cH)$ have $l>\sqrt{N}$, and are therefore not in the 
semi-classical IR sector. 
These are best described in terms of the {\em non-local} string states 
\begin{align}
 \psi_{x,y} := \ket{x}\bra{y} \qquad \in End(\cH)   
\end{align}
introduced in \cite{Iso:2000ew} and
discussed in detail below.
These form the core of the fuzzy or ''quantum`` geometry, yet they are often neglected.
The most extreme example on $S^2_N$ is the   state with maximal $J_3$ eigenvalue,
\begin{align}
 Y^{N-1}_{N-1} &= |p\rangle\langle -p|, 
\end{align}
where $|p\rangle$ is the highest weight state,
cf. \cite{Andronache:2015sxa}. This is in the far UV region of the 
algebra, it is maximally de-localized 
and has maximal angular momentum  $l_{UV}=2j \sim N \gg \sqrt{N} = l_{NC}$.

Since these string states comprise the bulk of the algebra of functions,
it should not be surprising that they  lead to significant non-local contributions 
in the effective action. The resulting string-like theory will be elaborated below.
In the context of quantum mechanics, the analogous types of de-localized
density matrices lead to the well-known non-local entanglement and EPR-type considerations,
which are characteristic for the ''deep quantum`` regime.
Clearly such states are not well-described 
by deformation quantization or in any semi-classical picture, yet they form the 
core of noncommutative (or fuzzy) field theory.

\paragraph{Rescaling and planar limit.}

So far, the radius of $S^2$ was fixed to be $R_N^2$. Now  introduce a scaling factor so that 
\begin{align}
  X^a X_a = R^2 \ 
\end{align}
with any desired radius $R$.
Then for $R=1$ one obtains the classical sphere as $N\to\infty$,
and the dispersion of the coherent states is
\begin{align}
 \Delta^2(|x\rangle) = \langle x|\sum_a (X^a - \langle X^a\rangle)^2|x\rangle = \frac 2{N+1} = O(\frac 1N) \ .
\end{align}
Hence the quantum cells become small as $N\to\infty$.
On the other hand if we scale the radius as 
\begin{align}
 R^2 = N\theta/2 , \qquad  \ \ \theta = const, \  N\to\infty,
\end{align}
the generators $X^1, X^2$ generate the Moyal-Weyl quantum plane $\R^2_\theta$
near the north pole \cite{Chu:2001xi}
\begin{align}
 [X^i,X^j] = i\theta \epsilon^{ij} \quad + \cO(\frac 1N) \ ,
\end{align}
dropping the $X^3$ generator. This is valid for states localized near the origin (i.e. the north pole).
We can then recover the standard coherent states on the  Moyal-Weyl plane
as $|x\rangle = U_x|0\rangle$ where $U_x=\exp(i\phi_i J^i)$
for $x^i = R\,\epsilon^{ij}\phi_j$, which gives 
\begin{align}
  \langle x'|x\rangle 
   &= \langle 0|e^{-i \phi' J}e^{i \phi J} |0\rangle 
    =  e^{-\frac{i}{2\theta} x^i\varepsilon_{ij} {x'}^j} e^{-\frac{|x-x'|^2}{4\theta}} \ .
\end{align}
Hence the overlap between coherent states is confined to regions of size $\theta$. This is
the scale of noncommutativity $L_{NC}$, which marks the boundary between the IR regime
and the  UV regime, as discussed above.

Even though we focus on the fuzzy sphere, the construction of coherent states 
goes through quite literally for any quantized coadjoint orbit such as $\C P^n_N$,
and also for the Moyal-Weyl quantum plane $\R^{2n}_\theta$.
We refer to \cite{Perelomov:1986tf,Hawkins:1997gj,Schneiderbauer:2016wub} for more details in these cases.

\subsection{String states}

Now consider some quantized fuzzy space 
(such as $S^2_N$, $\C P^n_N$, or even $\R^{2n}_\theta$),
with coherent states as above satisfying 
\begin{align}
 \langle  x| y\rangle = \frac 1{c_N}\, \tilde \d_N(x,y), \qquad  \ c_N =  \frac{\dim \cH}{\rm Vol \cM}, \qquad 
 \int_\cM dx\, \tilde  \d_N(x,y) | x\rangle  = |y\rangle \ .
 \label{inner-product-general}
\end{align}
 We then define the string states as 
\begin{align}
\left|^x_{y} \right) 
 &:= \psi_{x,y} := \ket{x}\bra{y} \qquad \in End(\cH)   \nn\\
\left(^x_{y} \right|
 &:= \psi_{x,y}^\dagger := \ket{y}\bra{x}
\end{align}
cf. \cite{Iso:2000ew}.
We also define the momentum operators acting on $End(\cH)$
\begin{align}
 \cP^a\, \cO &:= [X^a, \cO], \qquad \cO \in End(\cH)   \nn\\
 \Box\, \cO  &:=   \cP^a \cP_a \cO 
\end{align}
with expectation values 
\begin{align}
\left(^x_{y} \right|  \cP^a \left|^x_{y} \right)
 &=  \tr\, \psi_{y,x} [X^a,\psi_{x,y}] 
 = \vec{\bf x}(x) -  \vec{\bf x}(y) \nn\\
 \left(^x_{y} \right| \cP^a\cP_a \left|^x_{y} \right)
 &=  \tr\, \psi_{y,x} [X^a,[X_a,.]] \psi_{x,y} \ \ 
 = E_{xy} \ .
 \label{propagator-exp}
 \end{align}
 Here
 \begin{align}
 E_{xy} &= (\vec{\bf x}(x) -\vec{\bf x}(y))^2 +  \Delta^2_x + \Delta^2_y 
 \label{PP-strings}
\end{align}
is the energy of a string state  given by its length square
plus their intrinsic zero point energy,
in units of the noncommutativity scale (note that $\cP^a = [X^a,.] \sim \theta^{ab} \del_b$
has dimension length).
 $\Delta^2_{x,y}$ denotes the uncertainty at $x$ and $y$, respectively,
 which is simply $\D^2$ for homogeneous spaces.
In particular, the string states $\psi_{x,y}$ have ``matrix momentum'' 
$\cP = x -   y$. This 
is consistent with previous observations \cite{Bigatti:1999iz,Bergman:2000cw,Jiang:2001qa} in noncommutative field theory,
which now have a precise mathematical realization in terms 
of the  string states.

We will also need the general matrix elements 
\begin{align}
 \left(^x_{y} \right| \cP^a \cP_a \left|^{x'}_{y'} \right)
 &= \bra{x}X^a X^a \ket{x'} \langle y' | y\rangle  + \langle x | x'\rangle \bra{y'}X^a X^a \ket{y} 
 - 2  \langle x| X^a\ket{x'} \bra{y'} X_a |y\rangle \nn\\
  &\approx  E_{xy} \, \langle x | x'\rangle \langle y' | y\rangle \ 
  \label{propagator-general}
\end{align}
to a very good approximation. Note that this is nearly diagonal, 
and $E_{xy}$ is  bounded from below by the scale
of noncommutativity \eq{PP-strings}.

The remarkable feature of the string states is that they have good localization properties in 
{\em both} position and momentum.
This makes them very interesting and novel from a QFT point of view.
Even though (or rather because) they are typically in the UV regime
far from the semi-classical regime, 
they are very important for loop computations.

We also note that a non-commutative background in the matrix model can generally be 
viewed as a condensate of diagonal string states
\begin{align}
 X^a \sim \int d^n x x^a |x\rangle\langle x| \ .
\end{align}
This is exact for $\R^{2n}_\theta$ and for quantized homogeneous spaces, and holds at least approximately in general, 
cf. section \ref{sec:quasicoherent}.

\paragraph{Propagator.}

Generalizing the over-completeness relation \eq{inner-product-general}, we can write
\begin{align}
 c_N^2\int\limits_{\cM\times \cM}dx dy \left|^x_{y} \right) \left(^x_{y} \right| \ = \ \one_{End(\cH)} 
\end{align}
which follows again by group invariance.
In the same spirit, we can 
 state the central formula of this paper, which is an approximation 
for the propagator $(\Box + \mu^2)^{-1}$ using the string states. We claim that
\begin{align} \fbox{$
 (\bar \Box + \mu^2)^{-1} := c_N^2\, \int\limits_\cM dx dy \left|^x_{y} \right)
 \frac 1{E_{xy} + \mu^2} \left(^x_{y} \right| 
 \ \approx \ (\Box + \mu^2)^{-1}
 $}
 \label{prop-approx}
\end{align}
is an excellent approximation to the propagator.
Although no rigorous estimates will be given here, 
this formula can be justified  by the following computation 
\begin{align}
 (\tilde\Box+\mu^2)^{-1}(\Box+\mu^2) \left|^x_{y} \right) 
  &= c_N^2\, \int dx' dy' \left|^{x'}_{y'} \right) \frac 1{E_{x'y'}+ \mu^2} \left(^{x'}_{y'} \right|(\Box+\mu^2)\left|^x_{y} \right) \nn\\
  &\approx c_N^2\,\int dx' dy' \left|^{x'}_{y'} \right) \frac 1{E_{x'y'}+ \mu^2} (E_{xy} + \mu^2)  
  \langle x' | x\rangle \langle y| y'\rangle \nn\\
   &\approx \ \left|^x_{y} \right) 
\end{align}
using \eq{propagator-general}  and \eq{inner-product-general}.
The  approximation here comes from the variation of $E_{x,y}$
on scales of order $L_{NC}$, which is small since $E_{x,y} \geq 2L_{NC}^2$.
We therefore expect \eq{prop-approx} to be an excellent approximation, even for $\mu^2=0$, 
and there is no problem with any singularities\footnote{Note also that the phases of the coherent states 
cancel out, so there is no hidden phase ambiguity.}.
We will see explicitly that it works very well in the examples discussed below.
This justifies replacing $\Box^{-1}$ by $\tilde\Box^{-1}$,
which is the key proposal.

Finally, 
we remark that on  noncommutative branes, the above ``matrix momentum'' $\cP^a$ is only indirectly related
to the usual momentum. E.g. for a semi-classical scalar field we have
$\cP^a \phi = \theta^{ab} \del_b\phi$, which
leads to a non-trivial relation between the effective metric on a brane 
(``open string metric'') and the induced metric on the target space via $\theta^{ab}$ \cite{Steinacker:2010rh}.
This is responsible for some of the unusual features of noncommutative field theory.

\subsection{One-loop computations using coherent states}
\label{sec:loop-comp}

The 1-loop effective action can be expressed in terms of the trace of some operator $\cO$
acting on the space of wavefunctions. For the case of complex-valued scalar fields on a fuzzy space, 
this is the space $End(\cH)$ of operators on the underlying Hilbert space, where $\cH$ is an irreducible representation of $G$.
This trace can be written in terms of the string states as follows
\begin{align}
  \Tr_{End(\cH)} \cO 
 &= \frac{(\dim \cH)^2}{(\rm Vol \cM)^2}\,\int\limits_{\cM\times\cM} dx dy
  \left(^x_{y} \right| \cO \left|^x_{y} \right) \ .
 \label{trace-coherent-End}
\end{align}
This is an exact formula for any homogeneous quantum space of a (compact) Lie group $G$ with 
coherent states as discussed above. To prove it, it suffices to note that rhs of \eq{trace-coherent-End} 
is a functional which is invariant under $G_L \times G_R$, and 
by the uniqueness of the singlet in $End(\cH)$ it must be proportional to the trace.
Note that the  integral over $\cM\times \cM$ makes sense even though 
the spin states $\psi_{x y}$ form a non-trivial bundle
over $\cM\times\cM$, and there is no global section. However any phase factors cancel out in \eq{trace-coherent-End},
and it does not matter whether we integrate over the bundle $\cB\times\cB$ or over the base.

Now consider the case of hermitian fields  $\phi=\phi^\dagger \in End(\cH)$, which
are realized by the string states as follows  $e^{i\varphi}\psi_{x y} + e^{-i\varphi} \psi_{y x}$.
This suggests that the phase factors might lead to  non-trivial interference effects and we should 
integrate over the entire bundles $\cB\times \cB$. Nevertheless, these effects cancel 
and the trace over hermitian operators is simply $\frac 12 \times$ the trace over all operators. Thus
\begin{align}
 \Tr_{Herm(\cH)} \cO &= \frac 12\frac{(\dim \cH)^2}{(\rm Vol \cM)^2}\,\int\limits_{\cM\times\cM} dx dy 
  \left(^x_{y} \right| \cO \left|^x_{y} \right)
  \label{trace-coherent-End-hermit}
\end{align}
where $Herm(\cH)$ denotes the hermitian operators on $\cH$.
To evaluate the matrix elements, it is sometimes more transparent to write 
\begin{align}
 \left(^x_{y} \right| \cO \left|^x_{y} \right) = \tr\big(|y\rangle\langle x| \cO (|x\rangle\langle y|)\big) \ .
 \label{end-trace-rewrite}
\end{align}

Using the formalism of quasi-coherent states \cite{Schneiderbauer:2016wub} reviewed in section \ref{sec:quasicoherent}, 
the above formulas should hold
also on rather generic quantum spaces  to a very good approximation, as long as the operators 
$\cO$ are sufficiently ``local''. In the present paper, we will focus on the case of quantized coadjoint orbits 
for simplicity. 

As a warm-up, we compute the trace of the Laplacian on the fuzzy sphere $S^2_N$.
Using  \eq{trace-coherent-End} and \eq{propagator-exp} we obtain
\begin{align}
 \Tr_{End(\cH)} [X^a,[X_a,.]] &=
 \frac{N^2}{(\rm Vol S^2)^2}\,\int_{S^2\times S^2} dx dx' \tr(|x\rangle\langle x'|) \big(|X^{a}(x')- X^{a}(x)|^2 + 2\Delta^2\big) 
 (| x'\rangle\langle x|) \nn\\
 &=  \frac{N^2}{(\rm Vol S^2)^2}\,\int_{S^2\times S^2} dx dx' (|X^{a}(x')- X^{a}(x)|^2 + 2 \Delta^2)  \nn\\
  &=  \frac{N^2}{\rm Vol S^2}\,\int_{S^2} dx (|X^{a}(e)- X^{a}(x)|^2 + 2 \Delta^2) \ .
\end{align}
Here  $| .|$ is the Euclidean distance  in target space $\R^3$, 
and $e$ is an arbitrary point on $S^2$, and $\cH = \C^N$. We parametrize $S^2$ with the standard normalization $Vol(S^2)=  4\pi$.
Then $X^a(x) \in \R^3$ are functions on $S^2$ normalized as
\begin{align}
R_N^2 = X^a X_a = \frac 14(N^2-1) \ 
\end{align}
and recalling  $\D^2 \approx \frac{N}{2}$ \eq{Delta-S4} we obtain
\begin{align}
 \Tr_{End(\cH)} [X^a,[X_a,.]] &\approx
 \frac 14\frac{N^2(N^2-1)}{4\pi^2}\,\int_{S^2} dx (|e_3- x|^2 + O(\frac 1N)) \ .
\end{align}
where $x$ is now normalized to 1.
Evaluating  the integral 
\begin{align}
 \int_{S^2} |e_3- x|^2 &= 2\pi\int_0^\pi d\theta \sin\theta ((1-\cos\theta)^2+\sin^2\theta) 
 = 8\pi
\end{align}
results in
\begin{align}
 \Tr_{End(\cH)} [X^a,[X_a,.]]
  = \frac 12 N^2(N^2-1)\big(1+ O(\frac 1N)\big) \ .
\end{align}
This agrees very well with the exact result
\begin{align}
  \Tr_{End(\cH)} [X^a,[X_a,.]] &= \sum_{j=0}^{N-1} j(j+1)(2j+1) = \frac 12 (N^2-1) N^2 .
\end{align}
More generally, we can compute  for any smooth function $f$
\begin{align}
 \Tr_{End(\cH)} f(\Box) 
 &= \frac{N^2}{(\rm Vol S^2)^2}\,\int_{S^2} dx \int_{S^2} dy f(R_N^2|x-y|^2 + 2 \D^2) \nn\\
 &=  \frac{N^2}{\rm Vol S^2}\,\int_{S^2} dx f(R_N^2|e_3-x|^2 + 2 \D^2) \nn\\
 &=  2\pi\frac{N^2}{\rm Vol S^2}\int_0^\pi d\vartheta \sin\vartheta f(R_N^2(1-\cos\theta)^2+\sin^2\theta)+ 2 \D^2) \nn\\
 &=  \frac{N^2}{2} \int_{-1}^1 d u f(2R_N^2(1-u)+ 2 \D^2)  \nn\\
 &\approx  \int_{0}^N  dj\, 2 j f(j^2 + 2 \D^2) \approx \sum_{j=0}^{N-1} (2j+1) f\big(j(j+1) + 2 \D^2\big) \ \nn\\
 &= \Tr_{j_{\rm max} }f(\Box_g + 2 \D^2) \ .
\end{align}
Hence the result agrees well with the classical trace over $f(\Box_g + 2 \D^2)$ for any smooth function $f$ 
with UV cutoff $j_{\max} = N-1$, and the shift by  $2\D^2=N-1$ is negligible for $N \gg 1$.
Here $\Box_g$ denote the classical Laplacian on $S^2$.

Upon closer examination,
this computation is actually a bit strange: the  contribution of the integral comes from 
 non-classical,  UV regime with angular momenta 
$l^2 \geq \D^2 = O(N)$,
where we can neglect the shift by $2\D^2$. 
This is  the regime where one should in general 
{\em not} trust the semi-classical approximation,
and  this computation only works  because  the spectrum of the matrix
Laplacian $\Box$ coincides {\em exactly} with that of the classical Laplacian $\Box_g$, even in the far UV regime.
Thus even though the string states $|x\rangle\langle y|$ cannot be approximated by any classical functions,
they  allow to compute e.g.  the classical heat kernel expansions, as long as the operators 
under consideration (such as $\Box$) have the classical spectrum even in the UV regime.
We will see below that the method works also in other cases, but then the 
result does not always correspond to the naive semi-classical expectation.

\subsubsection{One-loop propagator on $S^2_N$}

\begin{figure}[h]
\begin{centering}
\includegraphics[width=0.35\textwidth]{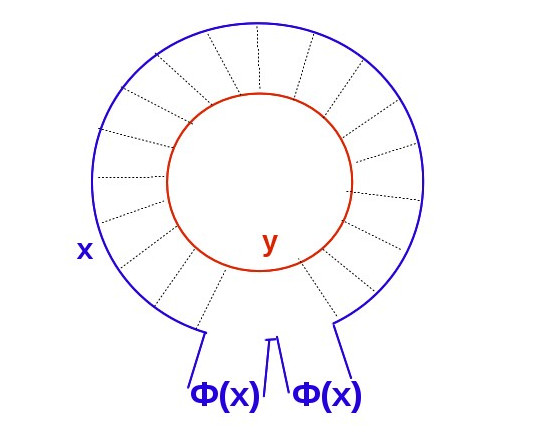}
\par\end{centering}
\caption{Planar 1-loop contribution.}
\label{fig:planar}
\end{figure}

As an application of this formalism, we want to compute the one-loop correction to the propagator 
for scalar $\phi^4$ theory on $S^2_N$, with 
hermitian scalar field $\phi^\dagger = \phi$ and action 
\begin{align}
 S[\phi] = \frac {1}{N} \tr\Big(\frac 12 \phi (\Box+\mu^2) \phi + \frac{g}{4!} \phi^4 \Big) \ 
  = S_0[\phi] + S_{\rm int}[\phi] \ .
 \label{S0-fuzzysphere}
\end{align}
The result will agree with the (more complicated and less transparent) original computation in \cite{Chu:2001xi}.
We use the standard normalization for $\Box$ 
\begin{align}
 X^a &= J^a_{(N)},  \qquad X^a X_a = \frac 14(N^2-1) = R_N^2  \nn\\
 \Box \phi &= [X^a,[X_a,\phi]]  \ 
\end{align}
with spectrum $l(l+1)$.
Then the effective action including  one-loop quantum corrections can be written as 
\begin{align}
 \Gamma_{\rm eff}[\phi] &= S[\phi] + \frac 12 \Tr_{End(\cH)} \log\Big(S''[\phi]\Big) \nn\\
 (\psi,S''[\phi]\psi) &= \frac {1}{N} \tr\Big(\psi(\Box+\mu^2)\psi + \frac g3 \phi^2 \psi^2
 + \frac g6 \psi\phi\psi\phi\Big)
\end{align}
where $S''[\phi]$ is the quadratic form for fluctuations around the background  $\phi$.
The one-loop contribution can be expanded follows 
\begin{align}
 \Gamma_{1-loop}[\phi] &= \Tr \log (.(\Box+\mu^2). + \frac g3 .\phi^2 . + \frac g6 .\phi . \phi ) \nn\\
 &=  \Tr \log (\Box+\mu^2) + \Tr\Big(.\frac 1{\Box+\mu^2}( \frac g3 \phi^2 . + \frac g6 \phi . \phi )\Big)
 +O(\phi^4) \ .
\end{align}
We assume that the background field
\begin{align}
 \phi =  \phi(X) \approx c_N\int\limits_{\cM} dy\,\phi(y)\ket{y}  \bra{y}
\end{align}
is slowly varying on the scale of noncommutativity.
Then 
$\phi$ acts nearly-diagonally 
on the string basis $\psi_{yx} = | y\rangle\langle x|$, and we can replace
\begin{align}
 \phi \psi_{yx} \approx \phi(y)\psi_{yx} \ 
\end{align}
and similarly for $\phi^2$.
Since $\dim\cH = N$,   \eq{trace-coherent-End} gives e.g. 
\begin{align}
 \Tr(.\phi^2 .) &= \frac {N^2}{Vol(\cM)^2}\int\limits_{\cM\times\cM} \! \!\! dx dy\tr (\psi_{y,x} \phi^2 \psi_{x,y}) \nn\\
  &= \frac{N^2}{ Vol(\cM)}\, \int\limits_{\cM} \! \!\! dx \, \langle x| \phi^2 |x\rangle \ .
\end{align}
Similarly, using the property \eq{propagator-exp} or \eq{prop-approx} of the propagator we find
\begin{align}
 Tr(.\Box^{-1}\phi^2 .) &= \frac {N^2}{Vol(\cM)^2}\int\limits_{\cM\times\cM} \! \!\! dx dy
 \tr (\psi_{y,x} (\Box+\mu^2)^{-1}(\phi^2 \psi_{x,y})) \nn\\
  &\approx \frac {N^2}{Vol(\cM)^2}\int\limits_{\cM\times\cM} \! \!\! dx dy
 \frac 1{R_N^2|x-y|^2 +2\Delta^2 + \mu^2} \tr (\psi_{y,x} \phi^2 \psi_{x,y}) \nn\\
 &= \frac{N^2}{Vol(\cM^2)}\, \int\limits_{\cM\times \cM} \! \!\! dx dy
  \frac 1{R_N^2|x-y|^2 + \tilde\mu^2}\langle x| \phi^2 |x\rangle \nn\\
  &= \frac{\mu_N^2}{Vol(\cM)}\, \int\limits_{\cM} \!  dx\,  \phi^2(x)
  \label{planar-comp-prop}
\end{align}
where
\begin{align}
 \tilde\mu^2 = \mu^2 + 2\Delta^2 > 0 
\end{align}
and $\mu^2_N$ is the 1-loop planar mass renormalization
\begin{align}
 \mu_N^2 &= \frac{N^2}{Vol(S^2)}\, \int\limits_{S^2} \! \!\! dy \,
  \frac 1{R_N^2|e-y|^2 + \tilde\mu^2}  \nn\\
 &= \frac{N^2}{2 R_N^2}\,\int_0^\pi d\vartheta\sin\vartheta
  \frac 1{(1-\cos\vartheta)^2+\sin\vartheta^2 + \frac{\tilde\mu^2}{R_N^2}} \nn\\
 &= 2\int_{-1}^1 du \frac 1{2-2u + \frac{\tilde\mu^2}{R_N^2}} \nn\\
 &\approx \sum_{j=0}^N  \frac{2j+1}{j(j+1) + \mu^2}
  =:  I^P
\end{align}
where $e$ is again some (arbitrary) reference point on $S^2$.
The approximation in  \eq{planar-comp-prop} consists of
replacing $\Box^{-1}$ by its diagonal matrix elements.
As discussed before (cf. \eq{prop-approx}),
this is justified as long as $\phi^2$ is in the IR regime, i.e. it varies only slowly at the NC scale.
Note also that $\Box^{-1}$ has bounded matrix elements in the string basis, which
ensures that there are no IR divergences in this integral.
This ``planar'' contribution is schematically depicted in figure \ref{fig:planar}.
It can be interpreted in terms of an open string from $x$ to $y$ propagating in the loop, integrated over $y$.

\begin{figure}[h]
\begin{centering}
\includegraphics[width=0.35\textwidth]{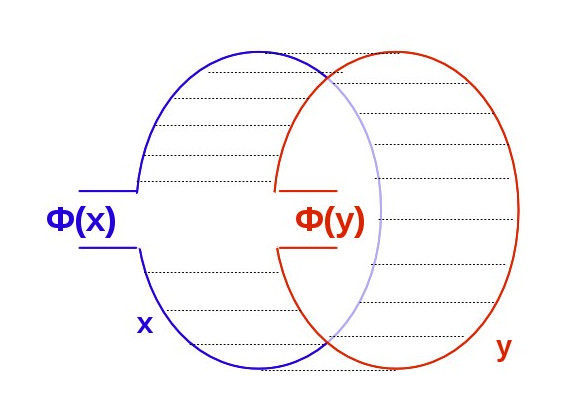}
\par\end{centering}
\caption{Non-planar 1-loop contribution.}
\label{fig:nonplanar}
\end{figure}

Now consider the ``non-planar'' contribution
 \begin{align}
 \Tr(.(\Box+\mu^2)^{-1}\phi . \phi)
 &= \frac {N^2}{ Vol(\cM)^2}\int\limits_{\cM\times \cM} \! \!\! dx dy \tr 
 (\psi_{y,x} (\Box+\mu^2)^{-1}(\phi \psi_{x,y} \phi)) \nn\\
&= \frac {N^2}{ Vol(\cM)^2}\int\limits_{\cM\times \cM} \! \!\! dx dy
 \langle x| (\Box+\mu^2)^{-1}\phi |x\rangle  \langle y|\phi |y\rangle \nn\\
 &= \frac {N^2}{ Vol(\cM)^2}\int\limits_{\cM\times \cM} \! \!\! dx dy
 \frac 1{R_N^2|x-y|^2 +\tilde\mu^2} \phi(x) \phi(y) 
\end{align}
depicted in figure \ref{fig:nonplanar}.
In contrast to the planar contribution
this results in a non-local term (!). This can be interpreted either in terms of an 
open string loop stretching from $x$ to $y$, or in terms of a  closed string propagating 
from $x$ to $y$.
Finally we compute the ``vacuum energy'' contribution 
\begin{align}
  \Tr \log (\Box+\mu^2) &= \frac{N^2}{(\rm Vol \cM)^2}\,\int_{\cM} dx \int_{\cM} dy 
  \log\big(R_N^2|x-y|^2 + \tilde\mu^2\big) \nn\\
  &= \frac{N^2}{4\pi}\,\int_{S^2} dx \log\big(|e-y|^2 + \frac{\tilde\mu^2}{R_N^2}\big) \nn\\
  &= \frac{N^2}2\int_{0}^2 du \log\big(2u + \frac{\tilde\mu^2}{R_N^2}\big) \nn\\
  &= N^2\Big(\ln 4-1 +  O(\frac{\tilde\mu^2}{R_N^2})\Big) \nn\\
  &=: \Gamma_{vac}
\end{align}
where $e$ is some point on the unit sphere $S^2$.
Then the one-loop contribution to the effective action up to quadratic order in $\phi$ is
\begin{align}\boxed{
\Gamma_{1-loop}  = \Gamma_{vac} + \frac g3 \frac{1}{\rm Vol(\cM)} \int\limits_{\cM} dx  \mu_N^2\phi(x)^2
   + \frac g6 \frac {N^2}{\rm Vol(\cM)^2 R_N^2}\int\limits_{\cM\times \cM} \! \!\! dx dy
 \frac {\phi(x) \phi(y) }{|x-y|^2+\frac{\tilde\mu^2}{R_N^2}} + O(\phi^4)
 }
 \label{1-loop-action-S2-closed}
\end{align}
cf.\footnote{The same structure was obtained in \cite{Iso:2000ew} using partitions into block-matrices. 
The present approach  is more efficient and does not require any 
ad-hoc partitions of space.} \cite{Iso:2000ew}.
 The planar contribution is local and  leads to a standard mass renormalization, 
 which agrees  with the results in \cite{Chu:2001xi} using a traditional mode expansion.
 We will see that the non-planar loop contribution also agrees with  \cite{Chu:2001xi}, but 
 it is now recognized as a long-range non-local   action. 
 This effect has no counterpart in standard quantum field theory. 
 It is of distinctly stringy nature\footnote{The present low-dimensional model should be viewed as non-critical 
 string theory. The connection to critical string theory will be discussed in section \ref{sec:one-loop}.},
 reflecting the presence of  virtual long strings described by the string states.
 Hence the model  describes a non-local theory
 even on scales much longer than the noncommutativity scale, and should not be considered as approximation to some local QFT.
 Although similar observation were made in \cite{VanRaamsdonk:2001jd,Kinar:2001yk},
 the present derivation based on string states is  most efficient, and easily generalized. 
 Clearly the higher loop contributions will add even more  non-local constrictions \cite{Minwalla:1999px},
 and could be obtained explicitly in a similar way
 (the extension to higher loops will be discussed briefly in section \ref{sec:general-pert-theory}). 

 The above derivation generalizes immediately to other, higher-dimensional fuzzy spaces such as 
 fuzzy $\C P^n_N$, noting that $|x-y|$ is always the Euclidean distance in target space.
 The one-loop effective action has always the same form  \eq{1-loop-action-S2-closed}, apart from  
 trivial adaptions. This is already a significant new result, 
 since non-planar loop contributions are very hard using group-theoretical  expansions
 and have not been performed.

 \paragraph{Comparison with 1-loop results for fuzzy $S^2_N$.}
 
 To check the validity of the approximations in the coherent state approach, we compare
 \eq{1-loop-action-S2-closed} with the known result for the fuzzy sphere \cite{Chu:2001xi}.
 We evaluate the non-planar contribution 
 for spin $l$ spherical harmonics $\phi = Y^l_m$, which gives 
\begin{align}
 \G_{NP} &= \frac g6 \frac{N^2}{(4\pi)^2R_N^2}\int\limits_{S^2\times S^2} \! dx dy
 \frac 1{|x-y|^2+\frac{\tilde\mu^2}{R_N^2}} \big(\sum_m (-1)^m Y^l_m(x)  Y^l_{-m}(y)\big) \nn\\
 &= \frac g6\frac{2l+1}{4\pi} \frac {N^2}{(4\pi)^2R_N^2}\int\limits_{S^2\times S^2} \! dx dy
 \frac 1{|x-y|^2+\frac{\tilde\mu^2}{R_N^2}} P_l(\cos\vartheta) \nn\\
  &= \frac g6\frac{2l+1}{2} \frac {N^2}{ R_N^2 (4\pi)} \int_0^\pi d\vartheta\sin\vartheta
 \frac{P_l(\cos\vartheta)}{(1-\cos\vartheta)^2+\sin\vartheta^2+\frac{\tilde\mu^2}{R_N^2}}  \nn\\
  &= \frac g6\frac {2l+1}{4\pi} \int_{-1}^1 du
 \frac{P_l(u)}{1-u+\frac{\tilde\mu^2}{2R_N^2}}  \nn\\
  &=:\frac g6 \frac {2l+1}{4\pi} I^{NP}(l)
  \label{nonplanar-fuzzyS2}
\end{align}
using the spherical harmonics addition theorem, where $\vartheta$ is the angle between $x$ and $y$.
This integral is convergent  as $R_N \to \infty$. 
Taking out the factor $(2l+1)$ from the sum over $m$, 
we recover precisely  the result 
which was obtained in \cite{Chu:2001xi} using a more complicated group-theoretical 
computation (which required the asymptotics of the $6J$ symbols). 
Clearly the present derivation is  much more efficient and transparent, 
it works equally well on higher-dimensional spaces such as $\C P^2$,
and - most importantly - it can be applied to more complicated problems such as 
 supersymmetric matrix models.

 \paragraph{Planar limit and UV/IR mixing on $\R^2_\theta$.}

 Although the above non-local term is perfectly well-defined on compact fuzzy spaces for finite $N$,
 it leads to IR-divergences in the non-compact limit $R\to\infty$, which clearly
 cannot be canceled by any local counterterms. This is  the 
 infamous UV/IR mixing of NC  field theory, 
 which is now understood in a completely transparent way. 
To see this, we recall that  the Moyal-Weyl quantum plane $\R^2_\theta$ can be obtained as a 
scaling limit of the fuzzy sphere (near the north pole) for 
$X^a = r J^a$ with $R^2 = r^2 R_N^2 = \frac{N\theta}4$ and fixed $\theta$.
Then the above non-planar contribution to the one-loop effective action takes the form
\begin{align}
 \G_{NP}  &\approx  \frac {gN^2}{6\rm Vol(\cM)^2}\int\limits_{\cM\times \cM} \! \!\! dx dy
   \frac{\phi(x)\phi(y)}{|x-y|^2+ \mu^2} \nn\\
   &=  \frac {g}{6\pi^2 \theta^2}\int\limits_{\cM\times \cM} \! \!\! dx dy
   \frac{\phi(x)\phi(y)}{|x-y|^2+ \mu^2} 
\end{align}
where ${\rm Vol} \cM = 4\pi R^2 = \pi N\theta$. Now $N$ has disappeared, and
this form can in fact be obtained directly\footnote{Starting with the noncompact case 
makes IR issues even more tricky,
while the present derivation is very clean.} from coherent state representation on $\cM =\R^2_\theta$.
Even though it is non-local,
this term is invariant under translations in the flat limit  $N \gg 1$, and 
we can compute it in a plane wave basis
$\phi(x) = \int \frac{d^2k}{2\pi} \phi_k(e^{i x k} + e^{-ikx})$.
This leads to 
 \begin{align}
 \G_{NP} &\approx\frac {g}{6\pi^2 \theta^2} \int d^2 k \phi(k)^2 \int  d^2 z
   \frac 1{|z|_g^2+\mu^2} e^{i k_i z^i} \nn\\
   &= \frac {g}{6\pi^2 \theta^2} \int d^2 k \phi(k)^2 \int  d^2 p
   \frac 1{p_i p_j G^{ij} +\mu^2} e^{i k_i \theta^{ij} p_j} .
   \label{UVIR-exp-standard}
 \end{align}
replacing $z^i = \theta^{ij} p_j$ in the second step.
Here $|.|_g$ is the background (closed string) metric, and 
 \begin{align}
 G^{ij} = \theta^{ii'}\theta^{jj'} \d_{i'j'}
 \end{align}
 is the ``open string'' metric which governs noncommutative field theory on $\R^2_\theta$ 
 \cite{Seiberg:1999vs,Steinacker:2010rh}.
This is the familiar form\footnote{In many papers on NC field theory 
\cite{Szabo:2001kg,Douglas:2001ba}, the kinetic term is 
defined as $\del_i \phi\del_i\phi$ rather than $[X^i,\phi][X_i,\phi]$. 
Then the closed string metric rather than the open string one appears in the last line of \eq{UVIR-exp-standard},
reconciling it with the literature. Conceptually, 
the present matrix model approach seems 
more natural.} for the non-planar contribution to the propagator 
on $\R^2_\theta$, and the derivation generalizes immediately to the case of $\R^{2n}_\theta$.
In this form, the non-locality leads to an IR divergence as $k\to 0$, 
and the well-known failure\footnote{This may be circumvented by adding additional terms to 
the action which strongly modify the noncommutative geometry, cf. \cite{Grosse:2012uv}.} 
of the standard renormalization procedure 
with local counterterms is  obvious given the non-local nature of the theory.
In particular, the  loop variables $p$ are now properly understood as position variables $x,y,z$.
The IR divergence in \eq{UVIR-exp-standard} 
suggests that the standard translation-invariant vacuum is inappropriate, and
the non-local equation of motion
 \begin{align}
 0 = \big(\Box + \mu^2 + \frac g3\mu_N^2\big)\phi(x) 
 + \frac g6 \frac {N^2}{\rm Vol(\cM) R_N^2} \int\limits_{\cM} dy\,\frac {\phi(y) }{|x-y|^2+\frac{\tilde\mu^2}{R_N^2}} 
 \end{align}
 suggest the presence of phase transitions and non-trivial ``striped'' vacua \cite{Gubser:2000cd,Bietenholz:2002vj,Panero:2016wwx}.
Analogous remarks apply to NC gauge theory.

 While this non-local nature of generic NC field theories excludes their application as fundamental theories, 
 they may still be  useful e.g. as effective description of  physics in  strong magnetic fields, 
 and possibly other contexts. 
 However, there is an important exception to this  conclusion,
 given by the  maximally supersymmetric IKKT and BFSS matrix models.
 We will see that  the nonlocality is much milder in the IKKT model, given by  10-dimensional  supergravity coupled to the brane. 
 More sophisticated backgrounds such as fuzzy $S^4_N$ in this model are promising candidates for the 
  quantum nature of space-time at short distances. The present methods are 
  also applicable in these backgrounds, as shown in \cite{Steinacker:2016yy}.

\subsection{(Quasi-) Coherent states on generic fuzzy spaces}
\label{sec:quasicoherent}

To show the applicability of the above coherent state methods to generic 
quantum geometries\footnote{We only have in mind here the case where the algebra of functions 
on a compact space is finite-dimensional. There are many examples where this is not satisfied,
and these are expected to have a very different intrinsic nature.},
we recall the general concept of quasi-coherent states introduced in \cite{Schneiderbauer:2016wub}, 
cf. \cite{Ishiki:2015saa,Berenstein:2012ts}. 
Given any background defined in terms of $D$ hermitean matrices $X^a \ \in End(\cH)$, 
they are 
defined to be the ground states  $|x\rangle$ of the point probe Hamiltonian
\begin{align}
 H_x = (X^a-x^a) (X_a-x_a), \qquad H_x |x\rangle= E(x)|x\rangle
\end{align}
for arbitrary $x\in \R^D$.
It follows that 
\begin{align}
 \langle H_{y}\rangle_{x}  & =\Delta^2(x)+|\vec{\bf x}(x)-\vec{y}|^{2} 
 \label{eq:lap_quadr_form}
\end{align}
where $\langle .\rangle_{x} = \langle x| . |x\rangle$ and $\vec{\bf x}(x) = \langle X^a\rangle_x$ and 
\begin{align}
\Delta^2(x) &:=\sum_{a=1}^{d} \langle \big(X^{a} - {\bf x^a}(x) \big)\big(X_{a} - {\bf x_a}(x) \big) \rangle_x
\label{eq:dispersion} 
\end{align}
is the dispersion.
We assume that this defines a ``brane'' i.e. a sub-variety  $\cM\subset \R^D$
where $\Delta^2(x)$ is small, and $E(x)$ grows quadratically in the  directions transversal to the brane.
We assume for simplicity that these ground states are non-degenerate,
defining a rank one projector 
\begin{align}
 |x\rangle\langle x| = P_0(H_x) .
\end{align}
Hence the $|x\rangle$ form a $U(1)$ bundle $\cB$ over $\cM$.
As in section \ref{coherent-fuzzy-S-2}, we can then map operators in $\phi\in End(\cH)$
to  functions via
\begin{align}
 \phi(x) = \bra{x} \phi \ket{x}
\end{align}
and 
\begin{align}
 \bra{x} [\phi,\psi]\ket{x} \approx i\{\phi,\psi\}
 \end{align}
defines a bracket on the classical functions which approximately satisfies the 
Leibniz rule and the Jacobi identity for large $N$. This recovers the Poisson bracket for functions.
The corresponding (NC) symplectomorphisms $U = e^{i\L(x)}$ define a connection $\nabla$ on $\cB$, 
whose curvature should be the symplectic form $\omega$ associated to the Poisson structure. 
In particular, the symplectic form will  satisfy the quantization condition 
\begin{align}
 \dim \cH  = {\rm Vol_\Omega(\cM)} = \int\limits_\cM \Omega 
\end{align}
where $\Omega = \frac 1{(2\pi)^n n!}\omega^{\wedge n}$ is the symplectic volume form on $\cM$.
We can then write down the following formula for a resolution of 
the unit in terms of the coherent states, generalizing \eq{coherent-states-S2}:
\begin{align}
 \one \ = \ \frac{\dim\cH}{\rm Vol \cM} \int\limits_\cM \Omega \ket{x}\bra{x} \ .
 \label{resolution-unity-general}
\end{align}
As a heuristic justification, we note that 
the expression on the rhs should be invariant under the connection $\nabla$ (since 
$\omega$ is the curvature of $\nabla$), and therefore invariant under symplectomorphisms.
This means that it should commute with the generators (at least to a very good approximation), 
and therefore it should be proportional to $\one$.
A rigorous proof or qualification of the  overcompleteness relation \eq{resolution-unity-general} in the generic case
is left as a challenge to future work.
As before, the  localization property of the coherent states can then be written as
\begin{align}
 \langle y|z\rangle &= \frac 1{c_N} \tilde \d(x,y), \qquad 
 c_N = \frac{\dim\cH}{\rm Vol(\cM)} \ .
\end{align}
If the above assumptions are satisfied, then all the formulae for the loop integrals 
developed in section \ref{sec:loop-comp}
are  applicable also in this general case.

\section{Higher loops and t'Hooft approach to NC QFT}
\label{sec:general-pert-theory}

Given these powerful  techniques, one would like to go beyond the one-loop approximation.
We will briefly discuss the generalization to higher loops in the spirit of t'Hooft's double line 
representation, and  possible non-perturbative setups.

Suppose we want to compute the  $n$-point functions  of a scalar field $\phi$ on some fuzzy space
at higher loops.
One approach  is
to first write down the perturbative contributions as usual using a Gaussian 
integration using some arbitrary but {\bf ordinary} basis for the matrix modes, and then to rewrite these 
Feynman rules in terms of the over-complete coherent state representations. 
This will result in a t'Hooft-type double line representation with  the 
simple propagators \eq{prop-approx}.
To make this more explicit, consider complex scalar fields on the fuzzy sphere 
with action $S[\phi] = S_0[\phi] + S_{\rm int}[\phi]$
with free part $S_0$ as in \eq{S0-fuzzysphere}, and  
$S_{\rm int}$ could contain any terms of the form $\frac 1N \tr (\phi^* \phi)^n$.
We can expand $\phi$ in an arbitrary basis
\begin{align}
 \phi^i_j = \sum_A \hat (Y^A)^i_j\, \varphi_{A}, \qquad   Y^{A} \in \cA= End(\cH)
\end{align}
where $i,j = 1,...,N$ labels a basis of $\cH$.
The correlators are obtained as usual from
\begin{align}
 Z[J] = \int_\cA D\phi e^{-S[\phi] + \tr \phi J} = e^{-W[J]} \ .
\end{align}
Then the perturbative expansion 
of a correlator is given be the sum of contractions.
This is most transparent in the matrix basis using $\phi^i_j$  rather than $\varphi_A$.
Then the free propagator 
\begin{align}
 \langle \phi^i_j\, {\phi^*}^k_l \rangle_0 \quad \in \cA \otimes \cA^*
\end{align}
is viewed as an  element in $\cA \otimes \cA^* \cong End(\cA)$, and represented by a double line  
starting at $({}^i_j)$ and ending at $({}^k_l)$. Since the vertices have the form 
of a matrix product, the Feynman rules are obtained directly in the t'Hooft double line organization,
where the labels $i,j$ of the lines are preserved in the vertices.
The diagrams are then be viewed naturally as ribbon graphs on a Riemann surface.
However, the labels are not preserved by the propagator, which makes the computations difficult.

The key is now to translate these Feynman-t'Hooft rules into the coherent state representation, 
for each given diagram. All we have to do is  use the form \eq{prop-approx} for the (approximate) 
propagator,
\begin{align}
 (\bar\Box + \mu^2)^{-1} =   c_N^2\, \int\limits_{\cM\times\cM} dx dy \left|^x_{y} \right) 
 \frac{1}{|x-y|^2+\tilde\mu^2} \left(^{x}_{y} \right| 
  \quad \in \cA \otimes \cA^*
\end{align}
which is explicitly written as an element in $\cA \otimes \cA^*$.
Since these propagators are connected by canonically contracting the indices i.e. 
evaluating $\cA = \cH\otimes \cH^*$ and $\cA^*= \cH^*\otimes \cH$, this gives immediately the 
Feynman-t'Hooft rules where the  lines of the propagator are now labeled by 
{\em positions} $x,y \in\cM$ on the fuzzy space which are preserved by the propagators,
and trivially connected at the vertices to form ribbon graphs.
The sums over the internal lines become position integrals over $\cM$.
The key feature is that both the propagators and the vertices are now diagonal in position space.
The resulting Feynman rules are very simple and natural, 
and their evaluation is much easier (!) than in ordinary QFT.
The one-loop diagrams lead to the diagrams \ref{fig:planar} and \ref{fig:nonplanar}, 
and the Feynman rules reproduce directly our results in section \ref{sec:loop-comp}, 
even quicker than using the trace-log forula.
It is then  easy to compute higher loop corrections;
some explicit computations will be presented  elsewhere \cite{Tekel:2016xx}
This simplification also leads to the hope that one may devise new techniques to extract their asymptotics, 
analogous to those in matrix models \cite{Brezin:1977sv}.

Since the resulting formalism is so simple, it is tempting to skip the intermediate steps and to 
use directly the coherent state representation of general operators 
$\phi\in End(\cH)$, e.g. as
\begin{align}
 \phi = \int_{\cM\times\cM}  dx dy |x\rangle \phi(x,y)\langle y| 
\end{align}
where $\phi$ is represented as a function on $\cM\times  \cM$.
Although this representation is not unique, one might 
try to define a path integral for such functions leading to the same type 2-line diagrams as before. 
On the other hand, it seems  more reasonably to replace the functions $\phi(x,y)$ by finite matrices 
\begin{align}
 \phi(x,y) \ \leftrightarrow \Phi_{x,y}
\end{align}
on some equidistributed  lattice on $\cM$ consisting of $\dim \cH$ points $x_i$, 
interpreted as (matrix) string with energy
\begin{align}
 \tilde\Box\Phi = [E,[E,\Phi]] = |x-y|^2 , \qquad E_{x,y} = x-y \ .
\end{align}
This is the picture of bi-local fields introduced in \cite{Iso:2000ew}.
Then the original model could be replaced by the simplified ``string'' matrix model 
\begin{align}
 S_{red}[\Phi] = tr([E,\Phi]^2 + \D^2 \Phi^2  + g \Phi^4)
\end{align}
which can be treated by the usual methods. 
Now the kinetic term is simplified,
and now represented by a single matrix $E$ instead of the set of matrices  $\{X^a\}$. 
In this form, non-perturbative approaches should be applicable.
A similar  simplification should apply for Yang-Mills matrix models.

The present techniques may also be useful in an exact RG approach to NC field theory,
noting that the modes with highest energy are the longest string modes.
In the case of a fuzzy sphere, these are the strings connecting opposite points, which 
should explain  the origin of the  antipodal terms found in \cite{Kawamoto:2015qla}.
We leave these topics for future investigations.

Finally, these ideas should also provide an efficient way to compute quantum corrections 
for {\em ordinary} $SU(N)$ Yang-Mills theory with large $N$ around 
non-trivial (Higgs) vacua corresponding to fuzzy 
extra dimensions \cite{Steinacker:2014lma,Aschieri:2006uw}.
Again the Feynman rules can be rewritten in a 
string basis for $\mmu(N)$ as above, and the propagators acquire weight factors 
corresponding to their distance in internal space as above.
Then the computations should be comparable to large $N$ gauge theory 
computations in the trivial vacuum.

\paragraph{Minkowski signature.}

The present paper is focused on the case of  Euclidean signature. 
In the case of Minkowski signature, $\Box$ has a 
non-trivial kernel, corresponding to time-like string states $\psi_{x,y}$ with  
\begin{align}
 (y-x)^2 + 2\Delta^2 +  \mu^2= 0
\end{align}
One might worry if such a model can ever be well-defined, 
but numerical simulations \cite{Kim:2011cr} demonstrate
that this can be achieved by adding suitable IR-regulator terms to the action.
We expect that the present techniques  provide a useful tool also in the 
case  of Minkowski signature, which is left for future work.

\section{The 1-loop effective potential for the IKKT model}
\label{sec:one-loop}

The above formalism is clearly also applicable to gauge theories, which are defined by 
matrix models of Yang-Mills type. In this section, we will use this to study
the one-loop effective actions for the maximally supersymmetric IKKT matrix model,
on some noncommutative brane background with the required properties as described above.
Again there is considerable overlap with \cite{Iso:2000ew}, but we develop a formalism applicable to generic fuzzy 
spaces, thus preparing the ground for the application on $S^4_N$ in \cite{Steinacker:2016yy}.
The background is defined in terms of 
10 hermitian matrices
\begin{align}
 X^a \sim x^a: \ \cM  \ \hookrightarrow \R^{10}
\end{align}
interpreted as quantized embedding function of some quantized symplectic manifold $\cM$ in $\R^{10}$.
They define the flux 
\begin{align}
 [X^a,X^b] = i\Theta^{ab} 
\end{align}
which corresponds to the quantized Poisson brackets of the $x^a$.
The IKKT or IIB matrix model is defined by the action
\begin{align}
 S_0[X] &= \frac 1{g^2}\Tr \Big(-[X_a,X_b][X^a,X^b]\, + 2\mu^2 X^a X_a \,  +  \obar\Psi \gamma_a[X^a,\Psi] \Big) \ .
 \label{bosonic-action}
\end{align}
To regularize possible IR singularities, 
we added a (small) mass $\mu^2$.
The equations of motion for the bosonic matrices are
\begin{align}
 (\Box + \frac 12 \mu^2) X_a  = 0 , \qquad \quad \Box = [X^a,[X_a,.]] \ .
\end{align}
Now consider fluctuations around some (not necessarily on-shell)
background $X^a \to X^a + \cA^a(X^a)$. Then the quadratic action for $\cA^a$ is given by
\begin{align}
 S[X+\cA]  &= S[X]  + \frac{2}{g^2}\Tr \Big(2\cA^a (\Box + \mu^2) X_a 
  +\cA_a \big((\Box + \mu^2)\d^a_b  + 2i [\Theta^{ab},. \, ] - [X^a,[X^b,.]]\big)  \cA_b\Big) \ . \nn
\end{align}
Hence the quadratic fluctuations $\cA^a$ are governed by the quadratic form
\begin{align}
\Tr \cA_a \Big((\Box + \mu^2)\d^a_b  + 2i [\Theta^{ab},. \, ] - [X^a,[X^b,.]]\Big) \cA_b  \ .
\label{fluct-action-nogf}
\end{align}
The last term can be canceled by adding a suitable 
Faddeev-Popov gauge-fixing term for $f= [\cA^a,X_a] =0$ \cite{Blaschke:2011qu}.
The one-loop  effective action  on a matrix background is defined by the Gaussian integration
around the background
\begin{align}
 Z[X] &= \int\limits_{\rm 1\, loop} d\cA d\Psi e^{-S[ X+\cA,\Psi]} 
 = e^{- \Gamma_{\!\textrm{eff}}[X] } \
 \end{align}
and we will denote the bare and one-loop contributions as
 \begin{align}
 \Gamma_{\!\textrm{eff}}[X] &= S_0[X] + \Gamma_{\!\textrm{1loop}}[X] \ .
\end{align}
We recall  
the following form of the one-loop effective action in the 
IKKT model \cite{Ishibashi:1996xs,Chepelev:1997av,Blaschke:2011qu} 
\begin{align}
\Gamma_{\!\textrm{1loop}}[X]\! &= \frac 12 \Tr \Big(\log(\Box + \mu^2 - M^{(\cA)}_{ab}[ \Theta^{ab},.])
-\frac 12 \log(\Box - M^{(\psi)}_{ab}[ \Theta^{ab},.])
- 2 \log (\Box)\Big)   \nn\\
 &= \frac 12 \Tr \Bigg(\sum_{n>0} \frac{1}n \Big((\Box^{-1}\big(-M^{(\cA)}_{ab}[ \Theta^{ab},.] 
    + \mu^2)\big)^n 
  \, -\frac 12 (-\Box^{-1}M^{(\psi)}_{ab}[ \Theta^{ab},.])^n \Big)  \Bigg) \nn\\
  &= \frac 12 \Tr \Bigg(\!\! \frac 14 (\Box^{-1}(M^{(\cA)}_{ab} [ \Theta^{ab},.] )^4 
  -\frac 18 (\Box^{-1}M^{(\psi)}_{ab} [ \Theta^{ab},.])^4 \,\, +  \cO(\Box^{-1}[ \Theta^{ab},.])^5 \! \Bigg) \nn\\
  &\quad + \frac 12 \mu^2 \Tr \Box^{-1} + O(\mu^4)
\label{Gamma-IKKT}
\end{align}
with  $a,b=1,...,10$,
where
\begin{align}
\begin{array}{rl}(M_{ab}^{(\psi)})^\a_\b &= \frac 1{4i} [\gamma_a,\gamma_b]^\a_\b \,  \\
                      (M_{ab}^{(\cA)})^c_d &= i(\d^c_b \d_{ad} - \d^c_a \d_{bd}) \, , \\
                    \end{array}  
\end{align}
and the  $2 \log \Box$ term arises from the ghost contribution.
Here $\Box$ and $ \Theta^{ab}$ refer to the operators defined for the background $X_i$ 
as in the previous sections.
Note that the coupling constant $g$ drops out  from $\Gamma_{\!\textrm{1loop}}$.
For $\mu=0$, the first non-vanishing term in this expansion is  $n=4$ due to maximal supersymmetry.
However for soft SUSY breaking with  $\mu^2\neq 0$, there are
contributions with $n=1$, starting with the above $\mu^2$ term.

This 4th order term plus the leading $\mu^2$ contribution 
is given by the following expression \cite{Blaschke:2011qu}:
\begin{align}
 \Gamma_{\!\textrm{1loop};4}[X]\! &=
 \frac 18 \Tr \Bigg(\!\!  (\Box^{-1}(M^{(\cA)}_{ab} [ \Theta^{ab},.])^4 
  -\frac 12 (\Box^{-1}M^{(\psi)}_{ab} [ \Theta^{ab},.])^4 \Bigg) \nn\\
 &= \frac 14 \Tr\Big(\Box^{-1}[\Theta^{a_1 b_1}, \ldots \Box^{-1}[\Theta^{a_4 b_4},.]]]]\Big) \nn\\
 &\quad  \big(-4 g_{b_1 a_2} g_{b_2 a_3} g_{b_3 a_4} g_{b_4 a_1} 
- 4 g_{b_1 a_2} g_{b_2 a_4} g_{b_4 a_3} g_{b_3 a_1} 
- 4 g_{b_1 a_3} g_{b_3 a_2} g_{b_2 a_4} g_{b_4 a_1} \nn\\
&\quad +  g_{b_1 a_2} g_{b_2 a_1} g_{b_3 a_4} g_{b_4 a_3}
 +  g_{b_1 a_3} g_{b_3 a_1} g_{b_2 a_4} g_{b_4 a_2}
+  g_{b_1 a_4} g_{b_4 a_1} g_{b_2 a_3} g_{b_3 a_2} \big) 
\label{Gamma-4-total}
\end{align}
and the leading term in $ \mu^2$ is
\begin{align}
 \Gamma_{\!\textrm{1loop};\mu^2}[X]\! &= - \frac 14 \mu^2 \Tr\big( \Box^{-1}\big) \ . 
\end{align}
To  explain the new technique for evaluating the trace using string states, 
we focus on the case of an irreducible fuzzy space of brane\footnote{The string theoretical picture is that 
 $N$ D-instantons  bound to and ``dissolved'' on a $D$-brane $\cM$.} given by the 
quantization of a symplectic manifold $\cM$; stacks of branes  will not be discussed here. 
We assume that there is an over-complete set of coherent states $|x\rangle$ on $\cM$,
with the associated string states $|y\rangle\langle x|$ spanning $End(\cH)$. 
According to the results in the previous sections, we can then write
\begin{align}
 \Box^{-1}(| x\rangle\langle y|) &\sim \frac 1{|x- y|^2 + 2\Delta^2} \ | x\rangle\langle y|  \nn\\
 \Box^{-1}[ \Theta^{ab},.] (| x\rangle\langle y|) 
 &\sim \frac 1{|x- y|^2+ 2\Delta^2}\, \d\Theta^{ab}( x,y) | x\rangle\langle y| 
 \end{align}
on a sufficiently slowly-varying background, where
 \begin{align}
 \d\Theta^{ab}(x,y) &:=  \Theta^{ab}(x)-  \Theta^{ab}(y) \ 
\end{align}
are now ordinary, commutative functions rather than operators.
 We assume that the dispersion 
 $\D_x^2 \approx \D^2$ is  independent of $x$ for simplicity.
Then the traces over $End(\cH)$ can be evaluated as
\begin{align}
\Gamma_{\!\textrm{1loop};4}[X]\! 
 &\sim \frac 14 \frac{(\dim\cH)^2}{(\rm Vol\cM)^2}\,
  \int\limits_{\cM\times\cM} \Omega_x \Omega_y  \frac{\d\Theta^{a_1b_1}(x,y)  \d\Theta^{a_2b_2}(x,y)
   \d\Theta^{a_3b_3}(x,y)  \d\Theta^{a_4b_4}(x,y)}{(|x- y|^2+2\Delta^2)^4}  \nn\\
 &\quad \ 3\big(-4 g_{b_1 a_2} g_{b_2 a_3} g_{b_3 a_4} g_{b_4 a_1} 
  + g_{b_1 a_2} g_{b_2 a_1} g_{b_3 a_4} g_{b_4 a_3} \Big)  \nn\\
  &= \frac 34 \, \int\limits_{\cM\times\cM} dx d y \rho(x)\rho(y)
   \frac{S_4[ \d\Theta(x,y)] }{(|x-y|^2+2\Delta^2)^4}  \nn\\
 \Gamma_{\!\textrm{1loop};\mu^2}[X]\! &\sim 
   \frac{5}2 \,\int\limits_{\cM\times\cM} dx d y \rho(x)\rho(y)
   \frac{\mu^2}{|x-y|^2+2\Delta^2} 
\label{1-loop-coh}
\end{align}
suppressing the target space metric $g_{ab}$. 
Here $\Omega_x = \rho(x) dx$ is the symplectic volume form on $\cM$ such that $\dim\cH = \rm Vol\cM$.
We denote accordingly
\begin{align}
 S_4[ \d\Theta] = -4tr \d\Theta^4 + (tr \d\Theta^2)^2 \ .
\end{align}
An important observation \cite{Tseytlin:1999dj} is the following:
If $ \d\Theta^{ab}(x,y)$ has  rank $\leq 4$, then 
\begin{align}
 - S_4[ \d\Theta] &= 
   4  tr( \d\Theta g  \d\Theta g  \d\Theta g  \d\Theta g) - (tr \d\Theta g  \d\Theta g)^2 \nn\\
  &=  4 ( \d\Theta_+^{ab}{ \d\Theta_+}_{ba})\, ( \d\Theta_-^{cd}{ \d\Theta_-}_{dc}), \qquad 
  \d\Theta_\pm =  \d\Theta \pm \star_g  \d\Theta \, \nn\\
   &\geq 0
\label{F+F-indentity}
\end{align}
where $\star_g$ denotes the 4-dimensional Hodge star with respect to  $g_{\mu\nu}$.
This leads to an attractive interaction, which vanishes
precisely in the (anti-) selfdual case $ \d\Theta = \pm \star_g \d\Theta$. 
Thus parallel $4$-dimensional branes with flux $\Theta^{ab}_A$ and $\Theta^{ab}_B$ are attracted to 
each other with an attractive $-\frac 1{r^4}$  potential \cite{Ishibashi:1996xs,Chepelev:1997av}
and are  unstable, 
unless $\Theta^{ab}_A - \Theta^{ab}_B$ is (anti-)selfdual.
For fluxes with rank $\geq 6$, the interaction is in general not attractive.
$\Gamma_{\!\textrm{1loop};4}$ vanishes identically for a single branes with
constant flux  such as $\R^4_\theta$, which reflects their BPS property.

For slowly varying backgrounds, $\Gamma_{\!\textrm{1loop};4}[X]$
describes  interactions which decay like $|x-y|^{-8}$, but are bounded 
for short distances by the NC cutoff $\Delta^2$.  
In the next section, we will identify  these interactions
with linearized IIB supergravity on  $\cM$,
generalizing previous results for block-matrix configurations and simple backgrounds 
\cite{Chepelev:1997av,Kabat:1997sa,Kabat:1997im,Chepelev:1998sm,Okawa:1998pz,Douglas:1998tk,Taylor:1998tv,Taylor:2001vb,Kitazawa:2002vh}.

As discussed in section \ref{sec:loop-comp},  possible
UV divergences are associated with large eigenvalues of $\Box$, 
which corresponds to widely separated points $x, y \in \cM$, or longs strings $|y\rangle\langle x|$. 
 This is the essence of UV/IR mixing.
 Due to the short-range interaction in the supersymmetric model, 
 this does not lead to any problems (at one loop)
 on manifolds with dimension less than 8, 
 in contrast to non-supersymmetric models.

\subsection{Induced interactions and linearized IIB supergravity}

Now we want to understand the  physics of the above one-loop interactions.
It was conjectured in \cite{Ishibashi:1996xs}  that the IKKT matrix model provides a non-perturbative definition
of IIB string theory  on $\R^{10}$. 
The main direct evidence  (i.e. based solely on the matrix model itself)
are loop computations as above for the interactions of simple branes in target space, which can be computed 
 in the matrix model and compared with string theory or rather IIB supergravity. 
The relevant (bosonic) degrees of freedom in IIB supergravity mediating such interactions
are the graviton,
the dilaton, and the anti-symmetric 2-form  and 4-form fields\footnote{The separation 
between the NSNS and the RR form fields is not clearly visible from the matrix model point of view.}. 
It was indeed found in \cite{Ishibashi:1996xs}, and corroborated in subsequent works 
\cite{Chepelev:1997av,Chepelev:1998sm,Okawa:1998pz,Douglas:1998tk,Taylor:2001vb,Kitazawa:2002vh}
that the interaction in the matrix model  matches with supergravity at least in the long-distance limit.
However, the methods were limited to  highly symmetric branes or 
 ``D-particles" represented by block-matrices. Using the above techniques, 
we can extend this to rather generic branes,
as long as they admit coherent states as discussed above.

First, consider the self-interaction of an irreducible brane background $\cM$. 
Expanding the above action using the short-hand notation 
\begin{align}
 \d\Theta(x,y) &= \Theta_x-\Theta_y  
 \end{align}
 we get 
  \begin{align}
 S_4[\d\Theta(x,y)] &= 
 - 4 tr\d\Theta(x,y)^4  + (tr\d\Theta(x,y)^2)^2 \nn\\
 &= -4 tr(T_x^2)  + tr(T_x)^2 + (x\leftrightarrow y)  \nn\\
 &\quad  +4\big(4 tr(\Theta_x\Theta_x \Theta_x \Theta_y)  - tr(\Theta_x \Theta_y)tr (\Theta_x\Theta_x) + (x\leftrightarrow y) \big) \nn\\
 &\quad - 16tr(T_x T_y )  + 2 tr T_x tr T_y \nn\\
 &\quad  - 8 tr(\Theta_x\Theta_y \Theta_x\Theta_y)  + 4 (tr(\Theta_x \Theta_y))^2 
 \label{S4-M}
\end{align}
which  disappears for $x=y$ as it must.
Here we identify the matrix-energy-momentum tensor of the (background) brane in target 
space as
\begin{align}
 T^{ab}[\Theta] &=  \Theta^{ac} \Theta^{cb} .
\end{align}
This is the ``closed string'' e-m tensor, in agreement\footnote{This should be contrasted with the effective (``open-string'') 
energy momentum tensor which arises in the 
effective gauge theory on the brane  with the open string metric, 
which has the standard form as in classical Yang-Mills gauge theory.} with related results in the literature, 
cf. \cite{Okawa:1998pz,Taylor:2001vb}. 
Furthermore, we denote the effective propagator on $\R^{10}$ as
\begin{align}
   D(x-y) = \frac{3}{2\pi^5} \frac 1{(|x-y|^2 + \D^2)^4} \sim \frac{3}{2\pi^5}\frac 1{|x-y|^8} \ ,
\end{align}
which for distances $|x-y|^2 \gg \D^2$ coincides with the 10-dimensional (Euclidean) propagator in 10 dimensions,
but is regularized in the UV by $\D^2$.
Then the effective interaction induced at one loop is
\begin{align}
 \Gamma_{\!\textrm{1loop};4}[X] 
 &\sim \frac{\pi^5}{2} \int\limits_{\cM\times\cM} dx d y \rho(x)\rho(y) 
  \Big( 2 S_4(\Theta(x)) \   D(x-y)  \nn\\
 &\quad  + 16 \big(\Theta^{ae}(x)\Theta_{ef}(x) \Theta^{f b}(x) 
 + \frac 14\Theta^{ab}(x) \Theta^{ef}(x)\Theta_{ef}(x)\big)\,  D^{(AS)}_{ab;cd}(x,y)\, \Theta^{dc}(y)  \nn\\
 &\quad - 8 T^{ab}(x) D^{(S)}_{ab;cd}(x,y)\, T^{cd}(y) \nn\\
 &\quad + 4 \Theta^{aab}(x)\Theta^{ef}(x) D^{(AS)}_{abef;cdgh}(x,y)\, \Theta^{cd}(y) \Theta^{gh}(y)  \Big) .
 \label{eff-IIB-interaction-loop}
\end{align}
Here 
\begin{align}
 D^{(S)}_{ab;cd}(x,y) &=  \big(g_{ac}g_{bd} + g_{ad} g_{bc} -\frac{1}{4} g_{ab} g_{cd}\big) D(x-y) \nn\\
 D^{(AS)}_{abef;cdgh}(x,y) &=  \big(g_{ac} g_{bd}  g_{eg} g_{fh}
     +g_{ac} g_{bh} g_{e d} g_{fg} - g_{ac} g_{bg} g_{e d} g_{fh}\big)D(x-y)  \nn\\
 D^{(AS)}_{ab;cd}(x,y)\ &= \big(g_{ac}g_{bd} - g_{ad} g_{bc}\big)D(x-y) 
\end{align}
For  $|x-y|^2 \gg \D^2$,
the third line  in \eq{eff-IIB-interaction-loop} can clearly be interpreted in terms of a
graviton exchange in $\R^{10}$, and $D^{(S)}_{\mu\nu;\a\b}(x,y)$ is indeed the graviton propagator in de Donder gauge.
The last line is  due to the exchange of a rank four antisymmetric tensor, and
$D^{(S)}_{\mu\nu;\a\b}(x,y)$ is  the propagator for a  rank four antisymmetric tensor.
The first line can be interpreted in terms of a dilaton exchange \cite{Taylor:1999pr,Kitazawa:2002vh}  
coupling the background density $\rho(y)$ to 
\begin{align}
 S_4(\Theta) =  -4 T^{ab}(x) T_{ab}(x) + T(x) T(x), \qquad T = T^{ab} g_{ab} \ .
\end{align}
Finally the second line can be interpreted as exchange of an antisymmetric rank 2 tensor field  $B_{ab}$, 
which couples to  branes\footnote{As explained in \cite{Taylor:1999pr},
the coupling of the brane to all the supergravity fields such as $B_{\mu\nu}$ etc. follows via T-duality 
from the well-known results that fundamental strings and lower-dimensional branes can be described
in terms of the field strength of the $U(N)$ gauge field in the world-volume of a $Dp$-brane.} 
via terms of the form \cite{Taylor:1999pr}
\begin{align}
 \int B_{ab} (\Theta^{ab} + \Theta^{ac}\Theta_{cd}\Theta^{db} + \frac 14 \Theta^{ab}\Theta_{cd}\Theta^{cd} + ...) .
\end{align}
Hence all these terms can be interpreted as interaction mediated by an exchange of 
the basic fields in 10-dimensional IIB supergravity, coupled
to a brane described by the matrix background $\Theta^{ab}$.
The specific form of the interaction ensures that it cancels identically 
for constant, flat backgrounds.
Even though this mechanism  has in principle been known for a long time \cite{Ishibashi:1996xs}, 
the derivations in the literature 
are based on separate block-matrices, and can be trusted only 
for large separations between localized branes. 
The present coherent state representation captures the detailed $x$-dependence for generic curved branes
within the matrix model.
With these  tools at hand, it should be possible to derive also the (analog of the)
Dirac-Born-Infeld effective action starting from the matrix model beyond the 
one-loop order,  
incorporating the full quantum effective action for slowly varying fields.

It is quite interesting that this interaction has a UV cutoff scale $\Delta^2$ in $D(x-y)$,
reflecting the quantum structure of the brane. This is as expected on branes 
with $B$-field, corresponding to noncommutative spaces. Moreover, 
the above derivation is easily adapted also to branes with vanishing 2-form flux, such as 
the fuzzy 4-sphere $S^4_N$ \cite{Castelino:1997rv}. This is properly understood as a degenerate 
higher-dimensional quantized symplectic space, where the $B$-field is 
averaged over the degenerate $S^2$-fiber over $S^4$
\cite{Ramgoolam:2001zx,Medina:2002pc,Steinacker:2015dra}. The present method allows to compute the one-loop 
effective action also on this background, in a much simpler and more transparent way than
 the group-theoretical approach in \cite{Steinacker:2015dra}.
This will be published elsewhere \cite{Steinacker:2016yy}.

\subsection{Fluctuations on a background}

Since the 1-loop interaction vanishes for flat backgrounds,  
the above action becomes more intuitive for
fluctuations $X^a = \bar X^a + \cA^a$ on some background brane $\cM$ described by $\bar X^a$. Then
\begin{align}
 \Theta^{ij} = \bar\Theta^{ij} + \cF^{ij}
\end{align}
where  $i\cF^{ij} = [X^i,\cA^j]- [X^i,\cA^j] +  [\cA^i,\cA^j]$ is an excitation on the background $\bar\Theta^{ij}$.
To organize the various contributions, we note again that $S_4$ depends 
only on the combination 
\begin{align}
  \d\Theta^{ab}( x,y) &:= (\bar\Theta^{ab}( x) - \bar\Theta^{ab}(y)) + (\cF^{ab}(x)) - \cF^{ab}(y))  \nn\\
   &=: \d\bar\Theta^{ab}(x,y) + \d\cF^{ab}(x,y) \ .
\end{align}
Assume that the background $\bar\Theta(x)$ is almost constant, 
while $\cF(x)$ is varying on much shorter scales (but still long compared to $\Delta^2$).
Then $\d\cF(x,y) \gg \d\bar\Theta(x,y)$, and 
we can organize $S_4(\cF)$ as follows
\begin{align}
 S_4(\d\Theta) &= S_4(\d\cF) + \cO(\d\cF^3 \d\bar\Theta)+ \cO( \d\cF^2 \d\bar\Theta^2)
  + \cO(\d\cF \d\bar\Theta^3) + S_4(\d\bar\Theta^4)
 \label{S4-expansion-F}
\end{align}
in decreasing order of significance.
We mainly focus on the leading $O(\d\cF^4)$ terms.
The mixed  terms describes
interactions of $\cF$ with the background flux $\bar\Theta$;
they may e.g. modify the propagator for the fluctuations $\cF$.
Finally  $S_4(\d\bar\Theta^4)$ corresponds to the self-interaction 
of the background as discussed in the previous section.

\paragraph{$O(\cF^4)$ term.}

The $O(\cF^4)$ term on a background $\cM$ has the same structure as the $S_4(\Theta)$ term, 
with the propagator  defined by the background $\cM$. 
It is most transparent for widely separated field configurations 
$\cF(x) = \cF_A(x) + \cF_B(x)$ where $\cF_{A,B}(x)$ have non-overlapping support. Then the interaction terms 
for such configurations have the by now familiar form 
\begin{align}
\Gamma_{\!\textrm{1loop};4}[\cF_A,\cF_B]\! 
&\sim \pi^5\int\limits_{\cM\times\cM} dx d y \rho(x)\rho(y) 
  \Big( 2 S_4(\cF_A(x)) \   D(x-y)  \nn\\
 &\quad  + 16 \big(\cF_A^{ae}(x)\cF_{Aef}(x) \cF_A^{f b}(x) 
 + \frac 14\cF_A^{ab}(x) \cF_A^{ef}(x)\cF_{Aef}(x)\big)\,  D^{(AS)}_{ab;cd}(x,y)\, \cF_B^{dc}(y)  \nn\\
 &\quad - 8 T_A^{ab}(x) D^{(S)}_{ab;cd}(x,y)\, T_B^{cd}(y) \nn\\
 &\quad + 4 \cF_A^{aab}(x)\cF_A^{ef}(x) D^{(AS)}_{abef;cdgh}(x,y)\, \cF_B^{cd}(y) \cF_B^{gh}(y)  \Big) .
\end{align}
There is a factor 2 which arises from the two possible associations $x\leftrightarrow A, y\leftrightarrow B$
and vice versa. 
This can again be interpreted as 
interactions due to the exchange of IIB sugra modes between the excitations $A$ and $B$.
Specifically, the first line is associated with a dilaton exchange with the background, 
the 2nd line  with the exchange of an antisymmetric rank 2 field, the 3rd line with a 
graviton exchange, and the last line  with exchange of a rank 4 tensor field.

We could also obtain a derivative expansion of the above interaction by 
expanding the $\d\cF = \cF(x)-\cF(y)$ into powers of $(x-y)$. Then the effective action for 
$\cF$ becomes a 4-th order derivative interaction with interaction strength given by $\D^{-4}$,
which was elaborated directly in \cite{Blaschke:2011qu}. 
Hence the above form provides a closed form for its long-distance behavior.
For the nonabelian case, this one-loop action is known to provide the leading 
$\cF^4$ term in an expansion of the DBI action
(cf. \cite{Blaschke:2011qu}), and the present technique should allow to corroborate this connection
in more detail.

\subsection{Non-supersymmetric matrix models}

Finally consider briefly the case of generic (non-supersymmetric) matrix models
and their relation with NC gauge theory.
As long as all fields are in the adjoint, the one-loop effective action can still be expressed in a similar way as \eq{Gamma-IKKT},
however starting at $O(\d\cF^2)$ rather than $O(\d\cF^4)$. 
At short distances, this  
leads to a derivative expansion starting with 2 derivatives of  $\cF$.
At long distances, the
propagators lead to a non-local interaction decaying like $(|x-y|^2 + \D^2)^{-2}$.

We can now make contact with the emergent gravity picture of NC gauge theory \cite{Steinacker:2010rh}:
The $U(1)$ sector of such a NC gauge theory defines
(in the local, semi-classical limit) a non-trivial effective (``open string'') metric for the remaining
fields. In accord with the  mechanism of induced gravity, the 1-loop integrals of any 
fields on such a background induces an
Einstein-Hilbert-type action in the effective action (among others).
In the case of NC field theory this arises due to IR modes in the loops
as verified in \cite{Blaschke:2010rr,Grosse:2008xr},
corresponding to the leading term in the above derivative expansion of $F$.
The new insights in the present paper complement this picture by an 
explicit form for  the induced long-distance interaction, which is due to the UV modes in the loops. 
In the case of maximal SUSY, this leads to 10D supergravity as shown above.
In generic non-SUSY models this interaction will in general not lead to
4-dimensional Einstein gravity, but  to a different type of shorter-range gravitational interaction.
However as shown in a companion paper \cite{Steinacker:2016yy}, 
the linearized 4D Einstein equations {\em do} emerge in the IKKT model, but only 
on  more sophisticated ``covariant'' noncommutative backgrounds and by a different mechanism.

\section{Conclusion}

One  message of this paper is that noncommutative field theory 
is very different from local field theory, 
and is more appropriately viewed as a theory of open strings ending on branes.
Although this insight is not new \cite{Seiberg:1999vs}, the formalism of bi-local  string states makes
this interpretation manifest and compelling from the noncommutative point of view.
The bulk of the kinematic phase space consists of an UV sector whose degrees of freedom 
are described by string states $|x\rangle\langle y| \in End(\cH)$, introduced previously in \cite{Iso:2000ew}. These are 
naturally interpreted as  open strings,
and behave completely differently from classical fields. 
We develop a formalism  based on integrals over string states  which
greatly simplifies the computation of the loop integrals.
This leads to a simple closed expression for the 
one-loop effective action in position space for generic fuzzy spaces, and
provides a clear picture of the non-locality encoded in the UV/IR mixing,
which arises from long strings with high energy.
The extension to higher loops is also indicated.
A rigorous proof or qualification of the overcompleteness relation \eq{resolution-unity-general} in the generic case 
is left as a challenge to future work.

In the maximally supersymmetric IKKT  matrix model,
the present formalism allows to derive directly the position space interactions which arise 
from quantum effects on fuzzy brane backgrounds, confirming the 
interpretation in terms of IIB supergravity.
This should provide an analytical tool to address the stabilization of 4-dimensional space-time
in the matrix model, cf. \cite{Steinacker:2015dra}.
It should also be possible now to derive directly the DBI action for branes in the matrix model.
Finally, the techniques developed here  are applied in \cite{Steinacker:2016yy} to the fuzzy 4-sphere,
which  exhibits 4-dimensional emergent gravity.

Even though generic non-commutative field theories defined by non-supersymmetric matrix models are non-local,
this does not exclude applications in suitable contexts such as condensed matter physics with strong magnetic fields.
Some of these models exhibit interesting phase structures 
\cite{Gubser:2000cd,Bietenholz:2002vj,Tekel:2015zga,DelgadilloBlando:2007vx,Steinacker:2005wj,Polychronakos:2013nca,O'Connor:2013rla,Panero:2016wwx},
and  the t'Hooft-like formalism proposed here should allow to greatly improve the analytic understanding of  
these models.
Furthermore, suitable limits of these models may  lead to non-trivial and interesting applications 
 \cite{Grosse:2012uv}. Therefore the  development of these powerful techniques 
should be useful also in these contexts.

\paragraph{Acknowledgements.}

I would like to thank H. Grosse and T. Koslowski  for  discussions, 
and J. Tekel for related collaboration.
This work was supported by the Austrian Science Fund (FWF) grant
 P28590 and by the Action MP1405 QSPACE from the European Cooperation in Science and Technology (COST).

\appendix

\bibliography{papers}
\bibliographystyle{diss}

\end{document}